# Exciton and interband optical transitions in hBN single crystal

L. Museur, [1] G. Brasse, [3] A. Pierret, [3,4] S. Maine, [3] B. Attal-Tretout, [5] F. Ducastelle, [3] A. Loiseau, [3] J. Barjon, [6] K. Watanabe, [7] T. Taniguchi, [7] and A. Kanaev [2]

[1] *Laboratoire de Physique des Lasers – LPL, CNRS UMR 7538, Institut Galilée, Université Paris 13, 93430 Villetaneuse, France*

[2] *Laboratoire d'Ingénierie des Matériaux et des Hautes Pressions – LIMHP, CNRS UPR 1311, Institut Galilée, Université Paris 13, 93430 Villetaneuse, France*

[3] *ONERA – Laboratoire d'Etude des Microstructures – LEM, ONERA-CNRS, UMR 104, BP 72, 92322 Châtillon Cedex, France*

[4] *CEA-CNRS Groupe ''Nanophysique et Semiconducteurs'', Institut Néel/CNRS, Université J. Fourier, CEA / INAC / SP2M, 17 rue des Martyrs, 38 054 Grenoble Cedex 9, France*

[5] *ONERA – Département Mesures Physiques – DMPh, 27 Chemin de la Hunière, 91761 Palaiseau Cedex, France*

[6] *Groupe d'Etudes de la Matière Condensée – GEMaC, Université de Versailles St Quentin, CNRS Bellevue, 1 Place Aristide Briand, 92195 Meudon Cedex, France*

[7] *National Institute for Materials Science, Namiki 1-1, Tsukuba, Ibaraki 305-0044, Japan*

**Abstract:**

Near band gap photoluminescence (PL) of hBN single crystal has been studied at cryogenic temperatures with synchrotron radiation excitation. The PL signal is dominated by the D-series previously assigned to excitons trapped on structural defects. A much weaker S-series of self-trapped excitons at 5.778 eV and 5.804 eV has been observed using time-window PL technique. The S-series excitation spectrum shows a strong peak at 6.02 eV, assigned to free exciton absorption. Complementary photoconductivity and PL measurements set the band gap transition energy to 6.4 eV and the Frenkel exciton binding energy larger than 380 meV.

Hexagonal boron nitride (hBN) is an anisotropic wide band gap semiconductor, constituted of graphite-like sheets with an hexagonal structure. It is nowadays one of the most promising materials for developing far UV light emitting devices.[1-3] Its electronic and optical properties have been the subject of many publications in the last decade. The understanding of electronic band and exciton structure of hBN is also a particularly important issue since it is considered as a reference system for BN nanotubes. [4-7] According to the most recent theoretical calculations, [8-10] hBN is an indirect band gap material, the electronic structure of which is governed by large Frenkel type excitonic effects. From the experimental point of view on the other hand, owing to its high-luminous efficiency comparable to that of pure ZnO single crystals, hBN is assumed to be a direct band gap material. [2, 3] Besides, the large dispersion in band-gap energy values reported earlier in literature, ranging from 4 eV to 7 eV, is currently explained as related to the sample quality and experimental methods used. In view of these particularities, a careful experimental analysis of near band gap optical properties of a high purity hBN single crystal is of particular interest. The luminescence properties of hBN have been earlier studied by cathodoluminescence [2, 11-15] or photoluminescence [11, 16-24] spectroscopy. In the present letter, we report on a detailed analysis of near band gap electronic properties of a high-purity hBN single crystal by means of photoluminescence (PL) and photoconductivity spectroscopy methods. From the comparison of PL and photoconductivity excitation spectra, we conclude about band gap and exciton binding energies of hBN.

The PL experiments were carried out at cryogenic temperatures with VUV synchrotron-radiation (SR) excitation of a pure hBN single crystal grown as described previously. [2, 25] Since simple deformation of a transparent crystal forms structural defects and affects the luminescence properties, [12, 15] the used crystal was carefully handled before to be fixed in the cryostat. The facility of the SUPERLUMI station (DESY, Hamburg) used in the experiments is explained in detail elsewhere.[26] Briefly, samples were cooled down to 10K and irradiated by monochromatized SR ($\Delta\lambda = 3.3 \overset{o}{A}$) under high vacuum (~$10^{-9}$ mbar). The measurements of luminescence spectra were carried out using a visible 0.275-m triple-grating ARC monochromator equipped with a CCD detector or a photomultiplier operating in the photon-counting mode. The pulse structure of SR (130 ps, 5 MHz repetition rate) enables time-resolved luminescence experiments at time-scale of 200 ns with a sub-nanosecond temporal resolution. Spectra can be recorded within a time gate $\Delta\tau$ possibly delayed with respect to the SR excitation pulse. The recorded spectra were corrected for the primary monochromator reflectivity and SR current.

The photoconductivity measurements were performed on high quality h-BN powders (Aldrich). The experimental setup is described in details in reference [27]. The microcrystals were dispersed in ethanol and then deposited on an appropriate electronic device with platinum interdigit electrodes. The solvent is evaporated at slow heating. The UV light, tuneable from 400 nm to 180 nm with a spectral resolution of 4 nm, is focused on the sample. A sourcemeter coupled with a preamplifier (Keithley 6430) is used to apply a constant bias voltage (4V on 2μm) on the electrodes in order to measure the photoinduced current on samples as a function of excitation wavelength. The spectrum was corrected by the apparatus function. More details will be given elsewhere.

The PL spectra of hBN single crystal excited with photon of energy 11.27 eV is displayed in Fig. 1. At room temperature a broad unstructured band is observed at about 5.4 eV. In contrast, this band shows a complicated fine structure at low temperature of 11K. Three lines labelled D4, D2 and D1 are observed at respectively 5.48 eV, 5.57 eV and 5.64 eV. A fourth line at 5.5 eV, labelled D3, can also be distinguished as a shoulder on the high energy side of line D4. These four emissions have been already reported and assigned to bound-exciton luminescence caused by disorder such as stacking faults or shearing of the lattice planes. [12, 15, 21, 28] Moreover, a broader emission, centred on 5.32 eV and previously assigned to quasi donor-acceptor transition (qDAP), [24] is also observed.

The PL spectrum presented in Fig. 1 is very similar to the spectrum previously measured with the same experimental set-up on commercial hBN powders.[24] Conversely, striking differences exist with respect to the laser induced PL spectrum of high purity hBN single crystal recently published [21]. The series of D lines assigned to bound excitons is observed in both spectra; on the other hand, the group of four sharp and intense S-lines in the energy range 5.7-5.9 eV reported in reference [21] and assigned to self-trapped exciton luminescence is not seen in Fig. 1. Apparently, this peculiarity cannot be explained by the sample quality. We believe that it is due to an extremely low intensity of SR light source. In fact, in the present experiment, the flux of photons on the sample is typically 0.5 ph/$\mu$m$^2$ per excitation pulse. In these conditions, the PL spectrum conveys the intrinsic energy relaxation in samples as independent excitation events free of cross-correlations. Under more intense lamp or laser excitation, the saturation of defects and traps usually favours defect-free luminescence. Interestingly, using excimer laser excitation (193 nm), we have also measured a PL spectra dominated by the S-lines series completely similar to the one reported in reference [21].

Indeed, a very weak S-series luminescence around 5.8 eV can be evidenced in the hBN single crystal, as shown in insert of Fig. 1. This PL spectrum was measured in the narrow time window $\Delta\tau = 0-2$ ns, which coincides with the SR excitation pulse. Such a choice is explained by the short excited-state lifetime of 0.6 ns [21] and allows us to increase considerably the signal-to-noise ratio. The lines observed at 5.778 eV and 5.804 eV can be respectively assigned to the exciton series S4 and S3 reported in ref. [[21]]. The two other exciton emissions S2 and S1 which have been observed previously were not seen in the present experiment but, as mentioned above, have been observed again in PL experiments with a laser excitation. Earlier reflectance spectroscopy measurements have shown that the S3 and S4 emissions result from the excitation of exciton states at 6.019 eV and 6.044 eV; they have been tentatively ascribed to the self trapped exciton (STE) luminescence. [21] The PLE spectroscopy can bring new important information about PL mechanisms and allows testing previous assignments.

The PLE spectrum of the S4 emission is displayed in Fig. 2. The sharp peak observed at 6.02 eV is followed by a plateau and then by a continuous increase of the intensity above about 6.4 eV. We assign the strong peak at 6.02 eV to the direct excitation of a free exciton: $hBN + h\nu_{exc} \rightarrow exciton$. Moreover the plateau observed in the energy range between 6.2 and 6.3 eV could be related to low-intensity excitonic transitions converging to the dissociation limit.[9] The large Stokes shift of the S4 emission, equal to 242 meV, results from the strong exciton-phonon interaction[29] and supports the assignment of the S4 line to the STE luminescence. Theoretical calculations[8-10] of the band structure and optical absorption spectra of hBN, taking into account electron-hole interaction, have underlined the importance of excitonic effects in this material. Compared to the peak observed at 6.02 eV, the strongest oscillator strengths for excitonic absorption are nevertheless calculated at somewhat smaller energies, ranging from 5.75 eV to 5.85 eV.

The layered structure of hBN could favour the STE formation. Two competing processes follow the excitation. In the first one the excited state lowers its energy by inducing a local distortion and tends to remain localized in this distorted site forming a STE. In the second process, the excited state lowers its energy by transfer from one site to a neighbour site forming so-called free exciton (FE). Generally, low coordination number of atoms in the crystalline lattice favours a formation of STE.[30] The hBN crystal is made of two-dimensional well separated layers, in which each atom is covalently bound to three neighbour ones. This low coordination number could explain the preferential formation of STE in hBN.

Interestingly, the PLE spectrum in Fig. 2 shows a monotonic increase in intensity for photon energies above 6.4 eV. We assign this feature to the exciton formation from free electron-hole pair following a band-to-band excitation $hBN + h\nu_{exc} \rightarrow e^- + h^+ \rightarrow exciton$. Our complementary photoconductivity measurements confirm this conclusion. The photoconductivity of hBN has been studied at room temperature under UV lamp excitation in the spectral range 230 – 190 nm. The induced photocurrent excitation spectrum is shown by open squares in Fig. 2. Two superimposed broad bands centred at 5.75 eV and 6 eV are observed. The photocurrent decreases for photon energies higher than 6 eV before increasing again above 6.3 eV. It is worth noting that no contribution to the photocurrent from the sharp FE absorption at 6.02 eV was detected. The photocurrent curve shown in Fig. 2 is consistent with that previously reported in hBN single crystal [31] and correlates well with recently published photostimulated luminescence (PSL) excitation spectrum [23]. The PSL arises from the trapping of free charge carriers in distant lattice sites and subsequent radiative recombination of carriers released from these traps by a visible light. [32] The two spectral features observed at 5.7 eV and 6 eV in the PSL excitation spectrum of hBN powder have been ascribed to the direct ionization of respectively acceptor and donor (or vice versa) levels [23] : $A^-(D^+) + h\nu_{exc} \rightarrow A^o(D^0) + e^-(h^+)$. We adopt this assignment to explain the two maxima observed 5.75 eV and 6 eV in the photocurrent excitation spectrum. Moreover, the photocurrent excitation curve fits very well with the PLE spectrum in the energy range above 6.2 eV. This correlation supports our assignment of the monotonic increase of PLE spectrum above 6.4 eV to the creation of free electron-hole pairs.

Finally, from a general point of view, as long as the hBN can be considered as optically thin material, the PLE spectrum reflects its intrinsic absorption. On the other hand, when the absorption becomes large non radiative recombinations, induced by surface defects, reduce the fluorescence quantum yield and lead to a saturation of the PLE signal. In principle, the shape of band edge absorption in hBN is governed by the 2D confinement effects as well as by the very small dispersion of conduction and valence bands in the MK direction [9, 10]. Accordingly, a sharp increase of the absorption should be observed at the band gap energy. Actually, when excitonic effects are taken into account, no clear step is expected at the threshold energy of the band edge absorption [10]. Therefore, we have assigned the inflexion observed at 6.4 eV on the PLE spectra (Fig. 2) to the onset of interband transitions and set the band gap to this energy namely $E_g$ = *6.4* eV. This value is slightly smaller than those obtained in the most recent theoretical calculations namely 6.47 eV [9], and 6.48 eV [10] for direct transitions. Actually, we have no clear evidence that our value corresponds to such direct transitions, so that the real direct band gap could be larger.

Similarly, our results suggest an excitonic binding energy larger than 380 meV, to be compared to the theoretical value of 700 meV. [9, 10] More accurate calculations of the absorption spectrum close to the edge of the continuum one-particle excitations are clearly needed for a better comparison between theory and experiment.

In conclusion, we have carefully analyzed the near band gap PL properties of the hBN single crystal with excitation by synchrotron radiation. The emissions due to qDAP and exciton D- and S- series have been observed. The D-series ascribed to the excitons trapped on stacking faults dominate the spectra; in the same time, the S-lines at 5.778 eV and 5.804 eV ascribed to the STE are found very weak. This is explained by an extremely low SR intensity, which allows considering elementary excitations as independent events free of cross-correlations. In these conditions far from any saturation effects, defects luminescence is favoured. The PLE spectrum of the 5.778 eV emission shows a strong peak at 6.02 eV, assigned to the free exciton absorption. Its high-energy spectral lineshape is inherent to the band gap absorption, which is confirmed by complementary photoconductivity excitation spectra measurements. The band gap energy of hBN 6.4 eV and Frenkel exciton binding energy larger than 380 meV have been obtained. These values are in general agreement with recent theoretical predictions.

Fruitful discussions with E. Gheeraert are gratefully acknowledged. This work has been supported by the project II-20080156 EC within the EU contract ELISA-226716, by the C'Nano IDF project Synthelas, as well as by the Eranet project, ANR-06-NSCI-006 and the ANR Cedona, ANR-07-NANO-007-02.

**References**

1. Watanabe, K., et al., *Far-ultraviolet plane-emission handheld device based on hexagonal boron nitride.* Nat Photon, 2009. **3**(10): p. 591.

2. Watanabe, K., T. Taniguchi, and H. Kanda, *Direct bangap properties and evidence for ultraviolet lasing of hexagonal boron nitride single crystal.* Nature Materials, 2004. **3**: p. 404-409.

3. Kubota, Y., et al., *Deep Ultraviolet Light-Emitting Hexagonal Boron Nitride Synthesized at Atmospheric Pressure.* Science, 2007. **317**(5840): p. 932-934.

4. Jaffrennou, P., et al., *Near-band-edge recombinations in multiwalled boron nitride nanotubes: Cathodoluminescence and photoluminescence spectroscopy measurements.* Physical Review B (Condensed Matter and Materials Physics), 2008. **77**(23): p. 235422.

5. Golberg, D., et al., *Boron Nitride Nanotubes.* Advanced Materials, 2007. **19**(18): p. 2413.

6. Arenal, R., X. Blase, and A. Loiseau, *Boron-nitride and boron-carbonitride nanotubes: synthesis, characterization and theory.* Advances in Physics, 2010. **59**(2): p. 101 - 179.

7. Wang, J., C.H. Lee, and Y.K. Yap, *Recent advancements in boron nitride nanotubes.* Nanoscale, 2010. **2**(10): p. 2028-2034.

8. Marini, A., *Ab Initio Finite-Temperature Excitons.* Physical Review Letters, 2008. **101**(10): p. 106405.

9. Arnaud, B., et al., *Huge excitonic effects in layered hexagonal boron nitride.* Physical Review letters, 2006. **96**: p. 026402.

10. Wirtz, L., et al., *Excitonic effects in optical absorption and electron-energy loss spectra of hexagonal boron nitride.* arXiv:cond-mat/0508421v1, 2005.

11. Silly, M.G., et al., *Luminescence properties of hexagonal boron nitride: Cathodoluminescence and photoluminescence spectroscopy measurements.* Physical Review B (Condensed Matter and Materials Physics), 2007. **75**(8): p. 085205.

12. Watanabe, K., et al., *Effects of deformation on band-edge luminescence of hexagonal boron nitride single crystals.* Applied Physics Letters, 2006. **89**(14): p. 141902.

13. Taylor, C.A., et al., *Observation of near-bandgap luminescence from boron nitride films.* Applied Physics Letters, 1994. **65**(10): p. 1251-1253.

14. Lukomskii, A.I., V.B. Shipilo, and L.M. Gameza, *Luminescence properties of graphite-like boron nitride.* Journal of applied spectroscopy, 1993. **57**(1-2): p. 607.

15. Jaffrennou, P., et al., *Origin of the excitonic recombinations in hexagonal boron nitride by spatially resolved cathodoluminescence spectroscopy.* Journal of Applied Physics, 2007. **102**(11): p. 116102.

16. Yao, B., et al., *Effects of degree of three dimensional order and Fe impurities on photoluminescence of boron nitride.* Journal of applied physics, 2004. **96**(4): p. 1947.

17. Era, K., F. Minami, and T. Kuzuba, *Fast luminescence from carbon-related defect of hexagonal boron nitride.* Journal of luminescence, 1981. **24/25**: p. 71.

18. Museur, L., et al., *Photoluminescence of hexagonal boron nitride: effect of surface oxidation under UV-laser irradiation.* Journal of Luminescence, 2007. **127**: p. 595-600.

19. Kanaev, A.V., et al., *Femtosecond and ultraviolet laser irradiation of graphite-like hexagonal boron nitride.* Journal of Applied Physics, 2004. **96**(8): p. 4483-4489.

20. Museur, L. and A. Kanaev, *Photoluminescence properties of pyrolytic boron nitride.* Journal of Materials Science, 2009. **44**(10): p. 2560-2565.

21. Watanabe, K. and T. Taniguchi, *Jahn-Teller effect on exciton states in hexagonal boron nitride single crystal.* Physical Review B (Condensed Matter and Materials Physics), 2009. **79**(19): p. 193104.

22. Watanabe, K., et al., *Time-resolved photoluminescence in band-edge region of hexagonal boron nitride single crystals.* Diamond and Related Materials, 2008. **17**(4-5): p. 830.

23. Museur, L., E. Feldbach, and A. Kanaev, *Defect-related photoluminescence of hexagonal boron nitride.* Physical Review B (Condensed Matter and Materials Physics), 2008. **78**(15): p. 155204.

24. Museur, L. and A.V. Kanaev, *Near band Gap photoluminescence properties of hexagonal boron nitride.* Journal of Applied Physics, 2008. **103**(10): p. 103520.

25. Taniguchi, T. and K. Watanabe, *Synthesis of high-purity boron nitride single crystals under high pressure by using Ba-BN solvent.* Journal of Crystal Growth, 2007. **303**(2): p. 525.

26. Zimmerer, G., *Status report on luminescence investigations with synchrotron radiation at HASYLAB.* Nuclear Instruments and Methods in Physics Research Section A: Accelerators, Spectrometers, Detectors and Associated Equipment, 1991. **308**(1-2): p. 178.

27. Brasse, G., et al., *Optoelectronic studies of boron nitride nanotubes and hexagonal boron nitride crystals by photoconductivity and photoluminescence spectroscopy experiments.* physica status solidi (b), 2010. **247**(11-12): p. 3076.

28. Watanabe, K., et al., *Band-edge luminescence of deformed hexagonal boron nitride single crystals.* Diamond And Related Materials, 2006. **15**(11-12): p. 1891-1893.

29. Toyozawa, Y., *Electron induced lattice relaxations and defect reactions.* Physica B+C, 1983. **116**(1-3): p. 7.

30. Toyozawa, Y., *Elementary processes in luminescence.* Journal of Luminescence, 1976. **12-13**: p. 13.

31. Remes, Z., et al., *The optical absorption and photoconductivity spectra of hexagonal boron nitride single crystals.* Physica Status Solidi A-Applications And Materials Science, 2005. **202**(11): p. 2229-2233.

32. Nagirnyi, V., et al. *Conduction band structure in oxyanionic crystals*. in *8th International Conference on Inorganic Scintillators and their use in Scientific and Industrial Applications SCINT 05*. 2005. Alushta, Crimea, Ukraine.

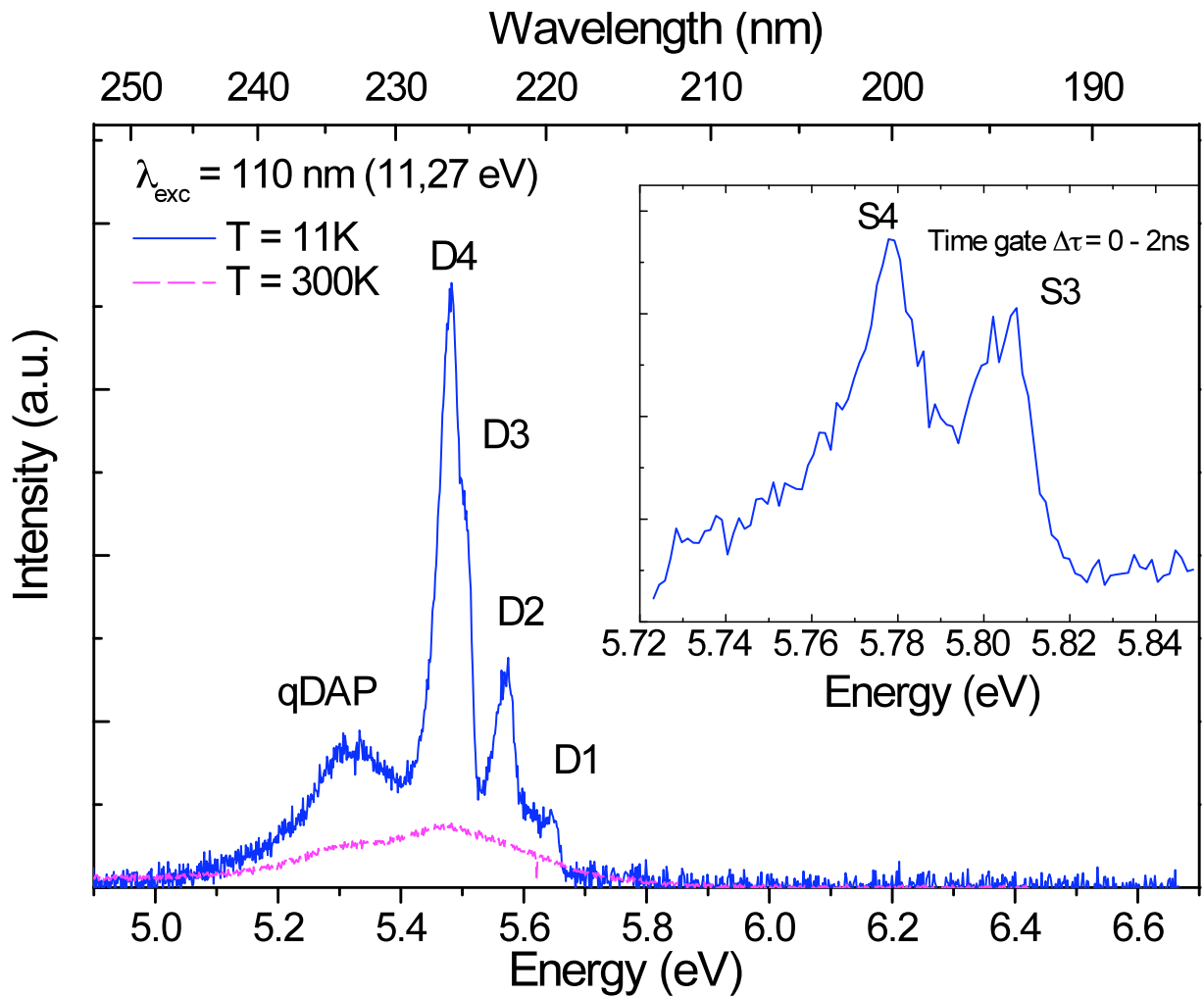

**Fig. 1** Time integrated PL spectra of hBN single crystal registered with an excitation energy of $E_{exc} = 11.27\,eV$. *Insert*: PL spectrum of hBN single crystal integrated on a time gate $\Delta\tau = 0-2\,ns$ with respect to the SR pulse (the excitation energy is always 11.27 eV)

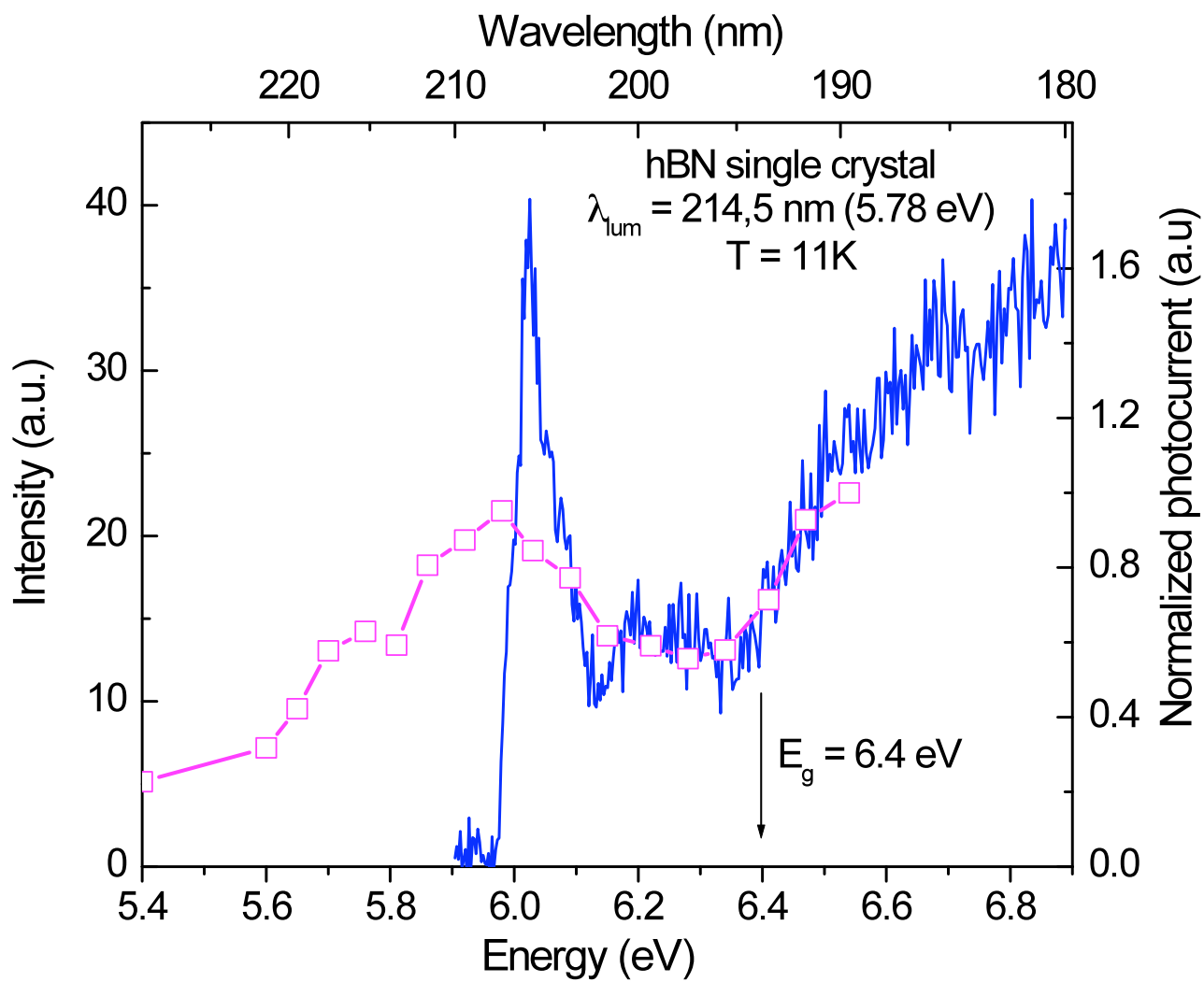

**Fig. 2** PLE spectra of the S4 emission (5.778 eV) in hBN single crystal at low temperature (straight line) and normalized photocurrent of hBN micro crystal at room temperature (open squares).